\documentclass[a4paper,10pt]{article}
\usepackage{amssymb}
\usepackage{amsthm}
\usepackage{amsmath,empheq}
\usepackage{mathrsfs}
\usepackage[dvipsnames]{xcolor}
\usepackage{tikz}
	\usetikzlibrary{decorations.pathreplacing}
\usepackage{pgfplots}

\usepackage[a4paper,top=1in, bottom=1.25in, left=1.2in, right=1.2in]{geometry}

\newtheorem{tm}{Theorem}
\newtheorem{lm}[tm]{Lemma}
\newtheorem{pr}[tm]{Proposition}

\newtheorem{con}[tm]{Conjecture}

\theoremstyle{definition}

\newtheorem*{ack}{Acknowledgements}

\theoremstyle{remark}

\newcommand{\dd}{\mathrm{d}}

\DeclareMathOperator{\dint}{\int\hspace{-5pt}\cdot\hspace{-2pt}\cdot\hspace{-5.5pt}\int\hspace{-5.5pt}}
\DeclareMathOperator{\idots}{\hspace{-2pt}\cdot\cdot}
\newcommand{\ballcol}{Mahogany}

\begin{document}
\title{On a kinetic equation in weak turbulence theory\\ for the nonlinear Schr\"odinger equation}%\\\ \\DRAFT}
\author{A.H.M.~Kierkels\\\small Institute for Applied Mathematics, University of Bonn\\\small Endenicher Allee 60, 53115 Bonn, Germany\\\small kierkels@iam.uni-bonn.de}
\date{June 2016}
{\linespread{1}
\maketitle
\begin{abstract}
The results from {\em J.~Stat.~Phys.~}{\bf159}:668-712 \& {\bf163}:1350-1393, on a quadratic kinetic equation in the analysis of the long time asymptotics of weak turbulence theory for the nonlinear Schr\"odinger equation, are summarized and placed in context. Additionally, two conjectures on self-similar solutions are presented, and backed with consistency analysis and numerics.\\


\end{abstract}

\section{Introduction}
The theory of weak turbulence, or wave turbulence\footnotemark[2], is a physical theory that aims to describe the transfer of energy between different spatial frequencies occurring in a large class of wave systems with weak nonlinearities. It was first used in \cite{P29} in the study of phonon interactions in anharmonic crystals, and the number of applications has increased over the years to include waves on fluid surfaces (e.g.~\cite{H62-63}, \cite{ZF66}, \cite{ZF67}), in plasmas (e.g.~\cite{Z67}, \cite{Z72}), in Bose-Einstein condensates (e.g.~\cite{ST95}, \cite{ST97}, \cite{S10}), in the early universe (cf.~\cite{MT03,MT04}), or on elastic plates (cf.~\cite{DJR06}). For a recent overview, containing a more exhausting list of examples and references, we refer the reader to \cite{N15}.\\
\footnotetext[2]{Depending on whom you ask, you will get a different answer to the question which term is to be used. Wave turbulence seems to be favoured by those who look more at physical applications of the theory. Moreover, in that context wave turbulence is usually presented as a state of being of a physical system. That is, a system exhibits wave turbulence precisely then when energy transfer between frequencies can be described by means of a kinetic equation. For the sake of remaining consist with previous works on the equations of interest, we continue to use the term weak turbulence.}

Starting point for any weak turbulence theory is a set of nonlinear wave equations, where the nonlinearity can be quantified by a real parameter $\varepsilon$. The objects of study are then the evolution equations for the moduli-squared in wave number space. However, for the sake of simplicity, let us just consider a translation-invariant wave equation for $u:\mathbb{R}\times\mathbb{R}^n\rightarrow\mathbb{C}$.

The linearized problem, obtained by setting $\varepsilon=0$, can be solved by using standard Fourier transform methods. Indeed, the space Fourier transform of the solution is $\hat{u}(t,k)=\hat{u}_0(k)e^{-i\omega t}$, where $\omega=\omega(k)$ is the dispersion relation. Moreover, since $\omega$ a real-valued function for conservative problems, the function $|\hat{u}|^2$ is then time-independent, and its evolution is trivial.

In the nonlinear case, i.e.~if $\varepsilon\neq0$, the evolution of $|\hat{u}|^2$ is nontrivial as a consequence of resonances between specific wave numbers $k$. Moreover, since the dynamics of $\hat{u}$ also depends on its phase, it is in principle not possible to obtain a closed equation for $|\hat{u}|^2$. However, weak turbulence theory argues that, for suitably chosen initial data, the evolution of $|\hat{u}|^2$ can approximated by a kinetic equation. In that case it is actually possible to give the evolution equation a particle interpretation.

Roughly speaking, we suppose our initial data to be of the form $\hat{u}_0=\sqrt\alpha\phi$, with $\alpha$ a nonnegative random variable, and $\phi$ a random phase. Moreover, we assume for the nonlinear terms that all functions are independent also for positive times. Averaging over phases and amplitudes we then expect to obtain a good approximation of the evolution of $|\hat{u}|^2$. For a more extensive road map from wave equations to weak turbulence equations, we refer the reader to Part II of \cite{N11}. However, the precise conditions under which this approach is valid have not been obtained. 

Note lastly that the derivation of kinetic equations in weak turbulence theory is not unlike the formal derivation of the Boltzmann equation from a particle system. In particular the assumption of continued statistical independence stands out.\\

In this paper we recall a formal derivation of the weak turbulence equation for the Schr\"odinger equation with a small defocusing cubic nonlinearity, in three space dimensions. From this equation we then derive an approximation to describe its long time behaviour, and we recall the main results on the quadratic equation at hand. We conclude by  posing two conjectures on the behaviour of self-similar solutions, which we back with consistency analysis and numerics.

\section{Weak turbulence theory for (NLS)}
One of the most widely studied equations in weak turbulence theory is the nonlinear Schr\"odinger equation:
\begin{equation}\tag{NLS}\label{eq:nls}
(i\partial_t+\Delta_x)u=\varepsilon|u|^2u,
\end{equation}
with $u=u(t,x):\mathbb{R}\times\mathbb{R}^3\rightarrow\mathbb{C}$ and $\varepsilon>0$ small. On a side note, recall that for suitable initial data the linear Schr\"odinger equation is explicitly solvable, where then the function $|\hat{u}|^2$ is time-independent, while solutions $u$ to \eqref{eq:nls} with $\varepsilon<0$ may exhibit blow-up of the $H^1$-norm in finite time (cf.~\cite{SS99}). In the following we present a formal derivation of the weak turbulence equation for \eqref{eq:nls}, where we follow the reasoning in \cite{N11}. The interested reader may also consult \cite{DNPZ92}, \cite{ZLF92}, or references therein.

Our aim is to derive an evolution equation for $n(\mathbf{k})=\langle|\hat{u}(\mathbf{k})|^2\rangle$, i.e.~the expected value of $|\hat{u}(\mathbf{k})|^2$ according to a probability distribution $\mathscr{P}_\mathbf{k}$ on the nonnegative real line, under the assumption that $\hat{u}(0,\cdot)$ is a field with random phases. Taking the space Fourier transform of \eqref{eq:nls}, we obtain
\begin{equation}\nonumber
(i\partial_t-|\mathbf{k}|^2)\hat{u}=\epsilon\left.\hat{u}\ast\hat{u}\ast\hat{\bar{u}}\right.\text{ with $\epsilon/\varepsilon$ constant},
\end{equation}
hence the function $\tilde{a}(t,\mathbf{k})=\hat{u}(t,\mathbf{k})e^{i|\mathbf{k}|^2t}$ satisfies
\begin{equation}\label{eq:wte_der_1}
i\dot{\tilde{a}}(\mathbf{k})=\epsilon\iint_{\left(\mathbb{R}^3\right)^2}\tilde{a}(\mathbf{k}_1)\tilde{a}(\mathbf{k}_2)\tilde{a}^*(\mathbf{k}_1+\mathbf{k}_2-\mathbf{k})e^{i(|\mathbf{k}_1+\mathbf{k}_2-\mathbf{k}|^2+|\mathbf{k}|^2-|\mathbf{k}_1|^2-|\mathbf{k}_2|^2)t}\dd\mathbf{k}_1\dd\mathbf{k}_2.
\end{equation}
However, the integrand in the right hand side of \eqref{eq:wte_der_1} is purely real on the submanifolds $\{\mathbf{k}_1=\mathbf{k}\}$ and $\{\mathbf{k}_2=\mathbf{k}\}$, and any contribution that results from the integral over either of them thus only affects the phase of $\tilde{a}$. As we are interested in the modulus $|\hat{u}|=|\tilde{a}|$, it thus makes sense to instead consider $a(t,\mathbf{k})=\hat{u}(t,\mathbf{k})e^{i|\mathbf{k}|^2t+2i\epsilon\int_0^t\|\hat{u}(s,\cdot)\|_{L^2(\mathbb{R}^3)}^2\dd s}$, which solves
\begin{equation}\label{eq:wte_der_2}
i\dot{a}(\mathbf{k})=\epsilon\iint_{\left(\mathbb{R}^3\right)^2}a(\mathbf{k}_1)a(\mathbf{k}_2)a^*(\mathbf{k}_1+\mathbf{k}_2-\mathbf{k})E_{\mathbf{k}_1\mathbf{k}_2}^{\mathbf{k}_1+\mathbf{k}_2-\mathbf{k},\mathbf{k}}(t)\dd\mathbf{k}_1\dd\mathbf{k}_2,
\end{equation}
with the shorthand
\begin{equation}\label{eq:wte_der_2b}
E_{\mathbf{r}_1\mathbf{r}_2}^{\mathbf{r}_3\mathbf{r}_4}(\tau)=\begin{cases}0&\text{ if $\mathbf{r}_1=\mathbf{r}_3$ \& $\mathbf{r}_2=\mathbf{r}_4$ or $\mathbf{r}_1=\mathbf{r}_4$ \& $\mathbf{r}_2=\mathbf{r}_3$,}\\e^{i(|\mathbf{r}_3|^2+|\mathbf{r}_4|^2-|\mathbf{r}_1|^2-|\mathbf{r}_2|^2)\tau}&\text{ else}.\end{cases}
\end{equation}
We now determine $a$ via a formal expansion in $\epsilon$ around a field $b$ with random phase; we set
\begin{equation}\label{eq:wte_der_3}
a(t,\mathbf{k})=b(\mathbf{k})+\epsilon a_1(t,\mathbf{k})+\epsilon^2 a_2(t,\mathbf{k})+\cdots,
\end{equation}
and recursively using \eqref{eq:wte_der_3} in \eqref{eq:wte_der_2} yields the expressions
\begin{equation}\label{eq:wte_der_4}
a_1(t,\mathbf{k})=-i\iint_{\left(\mathbb{R}^3\right)^2}b(\mathbf{k}_1)b(\mathbf{k}_2)b^*(\mathbf{k}_1+\mathbf{k}_2-\mathbf{k})\left.\int_0^tE_{\mathbf{k}_1\mathbf{k}_2}^{\mathbf{k}_1+\mathbf{k}_2-\mathbf{k},\mathbf{k}}(s)\dd s\right.\dd\mathbf{k}_1\dd\mathbf{k}_2,
\end{equation}
and, with abbreviated notation,
\begin{multline}\label{eq:wte_der_5}
a_2(t,\mathbf{k})=\dint_{\left(\mathbb{R}^3\right)^4}b_3b_4b_1^*b_2^*b_{1+2+k-3-4}\left.\int_0^t\int_0^sE_{1+2+k-3-4,3+4-k}^{12}(\sigma)\dd\sigma\,E_{34}^{3+4-k,k}(s)\dd s\right.\dd\mathbf{k}_1\idots\dd\mathbf{k}_4\\
-2\dint_{\left(\mathbb{R}^3\right)^4}b_1b_2b_{1+2-3}^*b_4b_{3+4-k}^*\left.\int_0^t\int_0^sE_{12}^{1+2-3,3}(\sigma)\dd\sigma\,E_{34}^{3+4-k,k}(s)\dd s\right.\dd\mathbf{k}_1\idots\dd\mathbf{k}_4.
\end{multline}
Further terms in the series \eqref{eq:wte_der_3} may be computed, but for the purpose of this formal derivation we may restrict ourselves to the ones given above.

As it turns out, in order to obtain an equation for $\mathscr{P}_\mathbf{k}$, it makes sense to consider its Laplace transform. This is the moment generating function $\mathscr{Z}_\mathbf{k}(\lambda)=\langle e^{\lambda|\hat{u}(\mathbf{k})|^2}\rangle$ for which, with \eqref{eq:wte_der_3}, we get
\begin{multline}\label{eq:wte_der_6}
\left\langle e^{\lambda|a(t,\mathbf{k})|^2}\right\rangle-\left\langle e^{\lambda|b(\mathbf{k})|^2}\right\rangle
=\left\langle e^{\lambda|b(\mathbf{k})+\epsilon a_1(t,\mathbf{k})+\epsilon^2 a_2(t,\mathbf{k})+\cdots|^2}-e^{\lambda|b(\mathbf{k})|^2}\right\rangle\\
=\left\langle e^{\lambda|b(\mathbf{k})|^2}\left(e^{\epsilon\lambda(b(\mathbf{k})a_1^*(t,\mathbf{k})+b^*(\mathbf{k})a_1(t,\mathbf{k}))+\epsilon^2\lambda(|a_1(t,\mathbf{k})|^2+b(\mathbf{k})a_2^*(t,\mathbf{k})+b^*(\mathbf{k})a_2(t,\mathbf{k}))+\cdots}-1\right)\right\rangle,
\end{multline}
where we can expand the term between round brackets in the right hand side of \eqref{eq:wte_der_6} as
\begin{multline}\nonumber
\epsilon\lambda\,2\Re\big(b^*(\mathbf{k})a_1(t,\mathbf{k})\big)
+\epsilon^2\lambda\left(|a_1(t,\mathbf{k})|^2+2\Re\big(b^*(\mathbf{k})a_2(t,\mathbf{k})\big)\right)
+\tfrac12\left(\epsilon\lambda\,2\Re\big(b^*(\mathbf{k})a_1(t,\mathbf{k})\big)\right)^2+\cdots\\
=\epsilon\lambda\,2\Re\big(b^*a_1\big)
+\epsilon^2\left((\lambda+\lambda^2|b|^2)|a_1|^2+\lambda\,2\Re\big(b^*a_2\big)
+\lambda^2\Re\big((b^*a_1)^2\big)\right)+O(\epsilon^3).
\end{multline}
Noting then that the phase averages $\langle b^*(\mathbf{k})a_1(t,\mathbf{k})\rangle_\phi$ and $\langle(b^*(\mathbf{k})a_1(t,\mathbf{k}))^2\rangle_\phi$ vanish, we find that \eqref{eq:wte_der_6} is approximated well by
\begin{equation}\label{eq:wte_der_7}
\mathscr{Z}_\mathbf{k}(t,\lambda)-\mathscr{Z}_\mathbf{k}(0,\lambda)
=\epsilon^2\left\langle e^{\lambda|b(\mathbf{k})|^2}\left((\lambda+\lambda^2|b(\mathbf{k})|^2)\left\langle|a_1(t,\mathbf{k})|^2\right\rangle_\phi+\lambda\,2\Re\left(\left\langle b^*(\mathbf{k})a_2(t,\mathbf{k})\right\rangle_\phi\right)\right)\right\rangle,
\end{equation}
while with \eqref{eq:wte_der_4}, and using \eqref{eq:wte_der_2b}, we immediately compute that
\begin{equation}\nonumber%\label{eq:wte_der_8}
\left\langle|a_1(t,\mathbf{k})|^2\right\rangle_\phi
=2\iint_{\left(\mathbb{R}^3\right)^2}|b(\mathbf{k}_1)|^2|b(\mathbf{k}_2)|^2|b(\mathbf{k}_1+\mathbf{k}_2-\mathbf{k})|^2\left|\int_0^tE_{\mathbf{k}_1\mathbf{k}_2}^{\mathbf{k}_1+\mathbf{k}_2-\mathbf{k},\mathbf{k}}(s)\dd s\right|^2\dd\mathbf{k}_1\dd\mathbf{k}_2.
\end{equation}
To compute the remaining phase average on the right hand side of \eqref{eq:wte_der_7} we now write $b^*(\mathbf{k})a_2(t,\mathbf{k})=I_1(t,\mathbf{k})-2I_2(t,\mathbf{k})$, where $I_1$ and $I_2$ are defined as the products of $b^*(\mathbf{k})$ and the first and second integral on the right hand side of \eqref{eq:wte_der_5} respectively. Exploiting then again \eqref{eq:wte_der_2b}, it is fairly straightfor\-ward to obtain
\begin{equation}\nonumber%\label{eq:wte_der_9}
\big\langle I_1(t,\mathbf{k})\big\rangle_\phi=2|b(\mathbf{k})|^2\iint_{\left(\mathbb{R}^3\right)^2}|b(\mathbf{k}_1)|^2|b(\mathbf{k}_2)|^2\left.\int_0^t\int_0^sE_{\mathbf{k}_1\mathbf{k}_2}^{\mathbf{k}_1+\mathbf{k}_2-\mathbf{k},\mathbf{k}}(s-\sigma)\dd\sigma\dd s\right.\dd\mathbf{k}_1\dd\mathbf{k}_2,
\end{equation}
and
\begin{multline}\nonumber%\label{eq:wte_der_10}
\big\langle I_2(t,\mathbf{k})\big\rangle_\phi=\frac12\times 2|b(\mathbf{k})|^2\iint_{\left(\mathbb{R}^3\right)^2}\left(|b(\mathbf{k}_1)|^2+|b(\mathbf{k}_2)|^2\right)|b(\mathbf{k}_1+\mathbf{k}_2-\mathbf{k})|^2\\\times\left.\int_0^t\int_0^sE_{\mathbf{k}_1\mathbf{k}_2}^{\mathbf{k}_1+\mathbf{k}_2-\mathbf{k},\mathbf{k}}(s-\sigma)\dd\sigma\dd s\right.\dd\mathbf{k}_1\dd\mathbf{k}_2.
\end{multline}
Thus, since for $\Omega\in\mathbb{R}$ we have that
\begin{equation}\nonumber%\label{eq:wte_der_11}
\frac1t\left|\int_0^te^{i\Omega s}\dd s\right|^2
=\frac1t\times2\Re\left(\int_0^t\int_0^se^{i\Omega(s-\sigma)}\dd\sigma\dd s\right)
=\tfrac1t\Big(\tfrac{2}{\Omega}\sin\big(\tfrac{\Omega t}{2}\big)\Big)^2\xrightarrow{t\rightarrow\infty}2\pi\delta(\Omega),
\end{equation}
we find that, for sufficiently large $t>0$, there approximately holds
\begin{equation}\label{eq:wte_der_12}
\tfrac1t\left(\mathscr{Z}_\mathbf{k}(t,\lambda)-\mathscr{Z}_\mathbf{k}(0,\lambda)\right)
=\lambda\eta(\mathbf{k})\left\langle e^{\lambda|b(\mathbf{k})|^2}\right\rangle+\left(\lambda^2\eta(\mathbf{k})+\lambda\gamma(\mathbf{k})\right)\left\langle|b(\mathbf{k})|^2e^{\lambda|b(\mathbf{k})|^2}\right\rangle,
\end{equation}
with
\begin{multline}\nonumber%\label{eq:wte_der_13}
\eta(\mathbf{k})=4\pi\epsilon^2\iiint_{\left(\mathbb{R}^3\right)^3}\left\langle|b(\mathbf{k}_1)|^2\right\rangle\left\langle|b(\mathbf{k}_2)|^2\right\rangle\left\langle|b(\mathbf{k}_3)|^2\right\rangle
\\\times\delta(|\mathbf{k}_3|^2+|\mathbf{k}|^2-|\mathbf{k}_1|^2-|\mathbf{k}_2|^2)\delta(\mathbf{k}_3+\mathbf{k}-\mathbf{k}_1-\mathbf{k}_2)\dd\mathbf{k}_1\dd\mathbf{k}_2\dd\mathbf{k}_3,
\end{multline}
and
\begin{multline}\nonumber%\label{eq:wte_der_14}
\gamma(\mathbf{k})=4\pi\epsilon^2\iiint_{\left(\mathbb{R}^3\right)^3}\left(\left\langle|b(\mathbf{k}_1)|^2\right\rangle\left\langle|b(\mathbf{k}_2)|^2\right\rangle-\Big(\left\langle|b(\mathbf{k}_1)|^2\right\rangle+\left\langle|b(\mathbf{k}_2)|^2\right\rangle\Big)\left\langle|b(\mathbf{k}_3)|^2\right\rangle\right)
\\\times\delta(|\mathbf{k}_3|^2+|\mathbf{k}|^2-|\mathbf{k}_1|^2-|\mathbf{k}_2|^2)\delta(\mathbf{k}_3+\mathbf{k}-\mathbf{k}_1-\mathbf{k}_2)\dd\mathbf{k}_1\dd\mathbf{k}_2\dd\mathbf{k}_3.
\end{multline}
However, the series for $a$ only significantly deviates from $b$ in times of order $\epsilon^{-1}$, which is large by the assumption that $\varepsilon>0$ is small. For times of order $\epsilon^{-1/2}$, we therefore consider $\langle|b(\mathbf{k})|^2\rangle$ and $\langle e^{\lambda|b(\mathbf{k})|^2}\rangle$ to be good approximations of $n(\mathbf{k})$ and $\mathscr{Z}_\mathbf{k}(\lambda)$ respectively, and $\frac1t(\mathscr{Z}_\mathbf{k}(t,\lambda)-\mathscr{Z}_\mathbf{k}(0,\lambda))$ of $\dot{\mathscr{Z}}_\mathbf{k}(\lambda)$, whereby from \eqref{eq:wte_der_12} we deduce
\begin{equation}\label{eq:wte_der_19}
\dot{\mathscr{Z}}_\mathbf{k}(\lambda)=\eta\left(\lambda+\lambda^2\partial_\lambda\right)\mathscr{Z}_\mathbf{k}(\lambda)+\gamma\left.\lambda\partial_\lambda\mathscr{Z}_\mathbf{k}(\lambda).\right.
\end{equation}
Taking the inverse Laplace transform of \eqref{eq:wte_der_19} then yields
\begin{equation}\nonumber%\label{eq:wte_der_20}
\dot{\mathscr{P}}_\mathbf{k}(s)=\eta\left.\partial_s\big(s\partial_s\mathscr{P}_\mathbf{k}(s)\big)\right.+\gamma\left.\partial_s\big(s\mathscr{P}_\mathbf{k}(s)\big),\right.
\end{equation}
and, computing the first moment, we finally arrive at the weak turbulence equation for \eqref{eq:nls}:
\begin{multline}\tag{WTE}\label{eq:wte}
\dot{n}(\mathbf{k})
=4\pi\epsilon^2\iiint_{\left(\mathbb{R}^3\right)^3}\Big(n(\mathbf{k}_1)n(\mathbf{k}_2)\big(n(\mathbf{k}_3)+n(\mathbf{k})\big)-\big(n(\mathbf{k}_1)+n(\mathbf{k}_2)\big)n(\mathbf{k}_3)n(\mathbf{k})\Big)
\\\times\delta(|\mathbf{k}_3|^2+|\mathbf{k}|^2-|\mathbf{k}_1|^2-|\mathbf{k}_2|^2)\delta(\mathbf{k}_3+\mathbf{k}-\mathbf{k}_1-\mathbf{k}_2)\dd\mathbf{k}_1\dd\mathbf{k}_2\dd\mathbf{k}_3.
\end{multline}
Note that even though \eqref{eq:wte} has been frequently studied (cf.~\cite{DNPZ92}, \cite{N11}, \cite{ZLF92}, or in the context of Bose-Einstein condensation \cite{JPR06}, \cite{LLPR01}, \cite{P92}, \cite{ST95,ST97}, \cite{S10}), its rigorous derivation is still a largely open question. However, see \cite{LS11} and \cite{LM15,LMN16} for first results in discrete NLS.

\section{The quadratic weak turbulence equation}
The paper \cite{EV15} presents an extensive study of isotropic solutions to \eqref{eq:wte}, i.e.~solutions to \eqref{eq:wte} that are of the form $n(\mathbf{k})=f(|\mathbf{k}|^2)$. Now, using this expression as an Ansatz in \eqref{eq:wte}, switching to spherical coordinates, and using the integral expression for $\delta$, it follows that $f$ should satisfy
\begin{multline}\nonumber
\dot{f}(k^2)
=4\pi\epsilon^2\iiint_{[0,\infty)^3}\Delta_{k_1k_2}^{k_3k}\Big(f(k_1^2)f(k_2^2)\big(f(k_3^2)+f(k^2)\big)-\big(f(k_1^2)+f(k_2^2)\big)f(k_3^2)f(k^2)\Big)\\\times\delta(k_3^2+k^2-k_1^2-k_2^2)\dd k_1\dd k_2\dd k_3,
\end{multline}
with
\begin{multline}\nonumber
\Delta_{k_1k_2}^{k_3k}
=k_1^2k_2^2k_3^2\times\iiint_{\left(\mathbb{S}^2\right)^3}\left[\frac1{(2\pi)^3}\int_{\mathbb{R}^3}e^{i\mathbf{s}\cdot(\mathbf{k}_3+\mathbf{k}-\mathbf{k}_1-\mathbf{k}_2)}\dd\mathbf{s}\right]\dd\Omega_1\dd\Omega_2\dd\Omega_3\\
%=k_1k_2k_3\times\frac1k\frac{(4\pi)^4}{(2\pi)^3}\int_0^\infty\sin(k_1s)\sin(k_2s)\sin(k_3s)\sin(ks)\frac{\dd s}{s^2}\\
=8k_1k_2k_3\times\frac{4\pi}{k}\int_0^\infty\sin(k_1s)\sin(k_2s)\sin(k_3s)\sin(ks)\frac{\dd s}{s^2},
\end{multline}
where the integral in the right hand side can be evaluated to $\frac\pi4\min\{k_1,k_2,k_3,k\}$ (cf.~\cite{ST97}). Thus, the isotropic version of \eqref{eq:wte} is given by
\begin{multline}\label{eq:wte_isot}
\dot{f}(\omega)
=4\pi^3\epsilon^2\iint_{[0,\infty)^2}\frac{K(\omega_1,\omega_2,\omega)}{\sqrt\omega}\Big(f(\omega_1)f(\omega_2)\big(f(\omega_1+\omega_2-\omega)+f(\omega)\big)\\
-\big(f(\omega_1)+f(\omega_2)\big)f(\omega_1+\omega_2-\omega)f(\omega)\Big)\dd\omega_1\dd\omega_2,
\end{multline}
with $K(\omega_1,\omega_2,\omega)=\min\{\sqrt{\omega_1},\sqrt{\omega_2},\sqrt{(\omega_1+\omega_2-\omega)_+},\sqrt{\omega}\}$. However, as the integral of $f$ is not conserved under the evolution \eqref{eq:wte_isot}, it is actually more convenient to study $g(\omega)=\sqrt\omega f(\omega)$, which after a suitable time rescaling satisfies
\begin{multline}\tag*{${\rm (WTE)}^\star$}\label{eq!:wte}
\dot{g}(\omega)=\frac12\iint_{[0,\infty)^2}K(\omega_1,\omega_2,\omega)\Bigg[\frac{g(\omega_1)}{\sqrt{\omega_1}}\frac{g(\omega_2)}{\sqrt{\omega_2}}\bigg(\frac{g(\omega_1+\omega_2-\omega)}{\sqrt{\omega_1+\omega_2-\omega}}+\frac{g(\omega)}{\sqrt{\omega}}\bigg)\\-\bigg(\frac{g(\omega_1)}{\sqrt{\omega_1}}+\frac{g(\omega_2)}{\sqrt{\omega_2}}\bigg)\frac{g(\omega_1+\omega_2-\omega)}{\sqrt{\omega_1+\omega_2-\omega}}\frac{g(\omega)}{\sqrt{\omega}}\Bigg]\dd\omega_1\dd\omega_2.
\end{multline}
This equation can be given several weak formulations, differing in the degree of interaction between the origin and the outer part of the solution (cf.~\cite{EV15}). From this point onwards we will restrict ourselves to weak solutions to \ref{eq!:wte} with fully interacting condensate, which are measure-valued functions that for test functions $\varphi\in C_c^2([0,\infty))$ satisfy
\begin{multline}\nonumber%\tag*{(WTE)$^{\rm w}$}\label{eq!:wte_w}
\partial_t\left[\int_{[0,\infty)}\varphi(\omega)g(t,\omega)\dd\omega\right]
=\frac12\iiint_{[0,\infty)^3}\frac{K(\omega_1,\omega_2,\omega_3)g(t,\omega_1)g(t,\omega_2)g(t,\omega_3)}{\sqrt{\omega_1\omega_2\omega_3}}\\
\times\big(\varphi(\omega_3)+\varphi(\omega_1+\omega_2-\omega_3)-\varphi(\omega_1)-\varphi(\omega_2)\big)\dd\omega_1\dd\omega_2\dd\omega_3.
\end{multline}

It was shown in \cite{EV15} that almost all weak solutions to \ref{eq!:wte} converge in the sense of measures to a Dirac mass at zero, while both mass (integral) and energy (first moment) are conserved. (Note that these quantities correspond to the conserved quantities $\|u\|_{L^2(\mathbb{R})}$ and $\|\nabla u\|_{L^2(\mathbb{R})}$ for solutions $u$ to the linear Schr\"odinger equation in $\mathbb{R}^3$, hence the terminology.) However, these weak limits have zero energy ($x\delta_0(x)\dd x\equiv0$). In order to investigate the disappearance of the energy, we suppose the long time behaviour of weak solutions to \ref{eq!:wte} to be well-approximated by a perturbation of a Dirac mass. To be precise, for long times we assume a solution $g$ to \ref{eq!:wte} with mass $1+\varepsilon$ to be of the form $g=\delta_0+G$, where $G$ is a nonnegative measure-valued function with mass $0<\varepsilon\ll1$ that satisfies a simpler equation. With the aim of determining a kinetic evolution equation for $G$, we consider the particle interpretation of the evolution of $g$ (cf.~Figure \ref{fig:wte}).
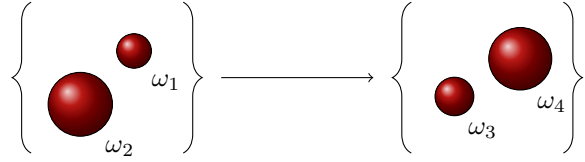
\begin{figure}[h]\center
\begin{tikzpicture}
\draw [color=white] (0,1) circle (.25cm);
\begin{scope}[xshift=-2.5cm]
\draw [decorate,decoration={brace,amplitude=.25cm}] (-1,-1) -- (-1,1);
\shadedraw [ball color=\ballcol] (sin{45}/2,cos{45}/2) circle (.228cm) node[label={[label distance=.078cm]300:$\omega_1$}]{};
\shadedraw [ball color=\ballcol] (sin{225}/2,cos{225}/2) circle (.423cm) node[label={[label distance=.273cm]300:$\omega_2$}]{};
\draw [decorate,decoration={brace,amplitude=.25cm}] (1,1) -- (1,-1);
\end{scope}
\draw [->] (-1,0) -- (1,0);
\begin{scope}[xshift=2.5cm]
\draw [decorate,decoration={brace,amplitude=.25cm}] (-1,-1) -- (-1,1);
\shadedraw [ball color=\ballcol] (sin{240}/2,cos{240}/2) circle (.255cm) node[label={[label distance=.105cm]285:$\omega_3$}]{};
\shadedraw [ball color=\ballcol] (sin{60}/2,cos{60}/2) circle (.414cm) node[label={[label distance=.264cm]285:$\omega_4$}]{};
\draw [decorate,decoration={brace,amplitude=.25cm}] (1,1) -- (1,-1);
\end{scope}
\end{tikzpicture}
\caption{The interaction mechanism in the particle interpretation of \ref{eq!:wte}. A pair $\{\omega_1,\omega_2\}\subset[0,\infty)$ interacts to form $\{\omega_3,\omega_4\}\subset[0,\infty)$ with $\omega_4=\omega_1+\omega_2-\omega_3$, where the rate of interaction is proportional to the ingoing particle density at particle sizes $\omega_1$, $\omega_2$ {\em and} $\omega_3$.}\label{fig:wte}
\end{figure}

Interactions where three or two of the particles $\omega_1,\omega_2,\omega_3\geq0$ are zero can readily be seen to be ``null-interactions'', i.e.~ones that have no effect on the distribution of particles. Indeed, using the fact that
\begin{multline}\nonumber
\varphi(\omega_3)+\varphi(\omega_1+\omega_2-\omega_3)-\varphi(\omega_1)-\varphi(\omega_2)\\
=(\omega_3-\omega_1)(\omega_3-\omega_2)\int_0^1\int_0^1\varphi''(\omega_1+\omega_2-\omega_3+s(\omega_3-\omega_1)+t(\omega_3-\omega_2))\dd s\dd t,
\end{multline}
it follows for any $\varphi\in C_c^2([0,\infty))$ that the mapping
\begin{equation}\nonumber
(\omega_1,\omega_2,\omega_3)\mapsto\frac{K(\omega_1,\omega_2,\omega_3)}{\sqrt{\omega_1\omega_2\omega_3}}\big(\varphi(\omega_3)+\varphi(\omega_1+\omega_2-\omega_3)-\varphi(\omega_1)-\varphi(\omega_2)\big),
\end{equation}
is continuous on $[0,\infty)^3$, and vanishes on the axes (cf.~\cite{EV15}, \cite{L13}) where the support of the product measure $(\delta_0\times\delta_0\times\delta_0)+(G\times\delta_0\times\delta_0)+(\delta_0\times G\times\delta_0)+(\delta_0\times\delta_0\times G)$ is found.

If only one of the particles $\omega_1,\omega_2,\omega_3\geq0$ is zero, the interaction does give a contribution. In the case where $\omega_3=0$, integrating out the Dirac mass yields
\begin{equation}\nonumber
\frac12\iint_{[0,\infty)^2}\frac{1}{\sqrt{\omega_1\omega_2}}\big(\varphi(0)+\varphi(\omega_1+\omega_2)-\varphi(\omega_1)-\varphi(\omega_2)\big)G(t,\omega_1)G(t,\omega_2)\dd\omega_1\dd\omega_2,
\end{equation}
while in either of the cases $\omega_1=0$ or $\omega_2=0$, we obtain
\begin{equation}\nonumber
\frac12\iint_{\{\omega_i>\omega_3\geq0\}}\frac{1}{\sqrt{\omega_i\omega_3}}\big(\varphi(\omega_3)+\varphi(\omega_i-\omega_3)-\varphi(0)-\varphi(\omega_i)\big)G(t,\omega_i)G(t,\omega_3)\dd\omega_i\dd\omega_3,
\end{equation}
and combining these integrals, we arrive at
\begin{multline}\nonumber
\partial_t\left[\int_{[0,\infty)}\varphi(x)G(t,x)\dd x\right]
=\frac12\iint_{\mathbb{R}_+^2}\frac{G(t,x)G(t,y)}{\sqrt{xy}}\big(\varphi(x+y)+\varphi(|x-y|)-2\varphi(x\vee y)\big)\dd x\dd y\\
+\frac12\iiint_{[0,\infty)^3}\frac{K(\omega_1,\omega_2,\omega_3)G(t,\omega_1)G(t,\omega_2)G(t,\omega_3)}{\sqrt{\omega_1\omega_2\omega_3}}\\\times\big(\varphi(\omega_3)+\varphi(\omega_1+\omega_2-\omega_3)-\varphi(\omega_1)-\varphi(\omega_2)\big)\dd\omega_1\dd\omega_2\dd\omega_3,
\end{multline}
where $x\vee y=\max\{x,y\}$. Normalizing then $G$ to be a probability measure, we find that the quadratic term is dominant for small $\varepsilon>0$, and we conclude that the function $G(t,\cdot)=\frac1\varepsilon G(\frac{t}\varepsilon,\cdot)$ to leading order satisfies
\begin{multline}\label{eq:qwte_weak}
\int_{[0,\infty)}\varphi(t,x)G(t,x)\dd x-\int_{[0,\infty)}\varphi(0,x)G(0,x)\dd x-\int_0^t\left.\int_{[0,\infty)}\varphi_s(s,x)G(s,x)\dd x\right.\dd s\\
=\int_0^t\left.\frac12\iint_{\mathbb{R}_+^2}\frac{G(s,x)G(s,y)}{\sqrt{xy}}\big(\varphi(s,x+y)+\varphi(s,|x-y|)-2\varphi(s,x\vee y)\big)\dd x\dd y\right.\dd s.
\end{multline}

It was noted in \cite{EV15}, under assumption of sufficient regularity and convergence of integrals, that \eqref{eq:qwte_weak} is the weak formulation of a kinetic equation of coagulation-fragmentation type. Indeed, rearranging the terms in their formulation, the quadratic approximation of the long time behaviour of solutions to \ref{eq!:wte} may be written as
\begin{multline}\tag{QWTE}\label{eq:qwte}
\partial_tG(x)
=\frac12\int_0^x\frac{G(x-y)G(y)}{\sqrt{(x-y)y}}\dd y
-\frac{G(x)}{\sqrt{x}}\int_0^\infty\frac{G(y)}{\sqrt{y}}\dd y\\
-\frac12\frac{G(x)}{\sqrt{x}}\int_0^x\left[\frac{G(x-y)}{\sqrt{x-y}}+\frac{G(y)}{\sqrt{y}}\right]\dd y
+\int_0^\infty\frac{G(x+y)}{\sqrt{x+y}}\left[\frac{G(x)}{\sqrt{x}}+\frac{G(y)}{\sqrt{y}}\right]\dd y,
\end{multline}
where the first two terms on the right hand side represent coagulation with a singular product kernel, and where in the last two terms one may recognise ``conditional fragmentation,'' in the sense that particles of size $x\geq0$ can break into particles of sizes $y\geq0$ and $x-y\geq0$ only if particles of either size already exist in the distribution. In fact, if we replace $(xy)^{-1/2}$ in \eqref{eq:qwte_weak} by any symmetric kernel, then formal calculations yield a strong formulation of this form.
\begin{figure}[h]\center
\begin{tikzpicture}
\draw [color=white] (0,1) circle (.25cm);
\begin{scope}[xshift=-5cm]
\draw [decorate,decoration={brace,amplitude=.25cm}] (-1,-1) -- (-1,1);
\shadedraw [ball color=\ballcol] (sin{300}/2,cos{300}/2) circle (.255cm) node[label={[label distance=.105cm]45:$x\wedge y$}]{};
\shadedraw [ball color=\ballcol] (sin{120}/2,cos{120}/2) circle (.379cm) node[label={[label distance=.229cm]225:$|x-y|$}]{};
\draw [decorate,decoration={brace,amplitude=.25cm}] (1,1) -- (1,-1);
\end{scope}
\draw [-] (-1.5,0) -- (-2.5,0) node[anchor=south]{$\mathbb{P}=\frac12$};
\draw [->] (-2.5,0) -- (-3.5,0);
\begin{scope}
\draw [decorate,decoration={brace,amplitude=.25cm}] (-1,-1) -- (-1,1);
\shadedraw [ball color=\ballcol] (sin{45}/2,cos{45}/2) circle (.255cm);
\shadedraw [ball color=\ballcol] (sin{225}/2,cos{225}/2) circle (.414cm);
\draw (0,0) node[label={[label distance=.2cm]120:$x$}]{};
\draw (0,0) node[label={[label distance=.2cm]330:$y$}]{};
\draw [decorate,decoration={brace,amplitude=.25cm}] (1,1) -- (1,-1);
\end{scope}
\draw [-] (1.5,0) -- (2.5,0) node[anchor=south]{$\mathbb{P}=\frac12$};
\draw [->] (2.5,0) -- (3.5,0);
\begin{scope}[xshift=5cm]
\draw [decorate,decoration={brace,amplitude=.25cm}] (-1,-1) -- (-1,1);
\shadedraw [ball color=\ballcol] (sin{300}/2,cos{300}/2) circle (.255cm) node[label={[label distance=.105cm]45:$x\wedge y$}]{};
\shadedraw [ball color=\ballcol] (sin{120}/2,cos{120}/2) circle (.444cm) node[label={[label distance=.294cm]225:$x+y$}]{};
\draw [decorate,decoration={brace,amplitude=.25cm}] (1,1) -- (1,-1);
\end{scope}
\end{tikzpicture}
\caption{One interpretation of \eqref{eq:qwte} as a particle system. Two particles $x,y\geq0$ interact at a rate that is proportional to the ingoing particle density at particle sizes $x$ and $y$, and with equal probability the largest particle is replaced by a particle of size $x+y$ or $|x-y|$. This is the sum of the processes of coagulation and conditional fragmentation, as described in the main text.}
\end{figure}
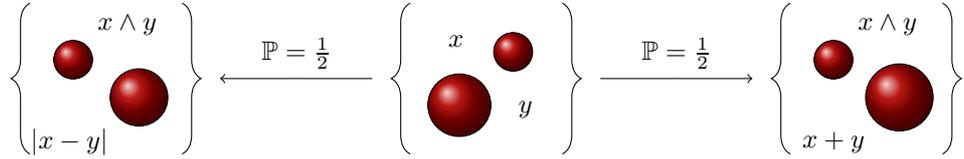

The paper \cite{EV15} further remarks on the similarity between \eqref{eq:qwte}, and equations that have been obtained by approximating the isotropic Boltzmann-Nordheim equation around a condensate to leading order, which contain additional terms resulting from quadratic terms in the Nordheim equation (cf.~\cite{DNPZ92}, \cite{LLPR01}, \cite{ST95,ST97}, \cite{S10}). Actually, in those papers systems of equations describing the evolution of the condensate and the non-condensed part of the solution were obtained, whereas our approximation does not explicitly assume a continued change of the Dirac mass at zero. 

Since \eqref{eq:qwte} was obtained in the study of transfer of energy towards infinity, this transfer is expected to continue in its evolution. The analyses in \cite{DNPZ92} and \cite{P92} are of interest in this study, and in particular \cite{P92} provides dimensional arguments for rescaling laws. In the same context, it was claimed in \cite{EV15} that there exist so called thermal equilibria for \eqref{eq:qwte}, which are solutions of the form $G(t,x)=cx^{-1/2}$ with $c\geq0$. However, although for $x>0$ the right hand side of \eqref{eq:qwte} then indeed vanishes, it can be checked that \eqref{eq:qwte_weak} with $G$ is of this form is not satisfied for test functions $\varphi$ with $\varphi(0)\neq0$. We do still expect inverse square root behaviour in the self-similar variable near zero in scaling solutions (cf.~Conjecture \ref{con:3}).

The rigorous study of \eqref{eq:qwte} was initiated in \cite{KV15}. With weak solutions defined to be continuous nonnegative measure-valued functions that satisfy \eqref{eq:qwte_weak} for suitable test functions $\varphi$, the first results from that paper can be reformulated as
\begin{tm}
Given a finite measure $G_0\geq0$, there exists at least one weak solution $G$ to \eqref{eq:qwte} with $G(0,\cdot)=G_0$. Moreover, any weak solution i) conserves mass and energy, ii) converges in the sense of measures to a Dirac mass at zero, and iii) has a strictly increasing mass at the origin.
\end{tm}\noindent
Again we thus find that functions with nonzero energies converge to a solution with no energy. In \cite{EV15} it had already been conjectured that energy might escape to infinity in a self-similar manner, yet in general there is no such thing as a scaling-invariant solution with multiple conserved moments, e.g.~mass and energy. However, the first moment of a function is independent of its value at zero, which paves the way to a modified notion of self-similarity.

\section{Self-similar solutions to \eqref{eq:qwte}}
It was noted in the introduction of \cite{KV15} that a weak solution $G$ to \eqref{eq:qwte} gives rise to the two-\linebreak[0]parameter family $\{G_{\kappa,\lambda}\}_{\kappa,\lambda>0}$ of weak solutions that are defined to be such that
\begin{equation}\nonumber
\int_{[0,\infty)}\psi(x)G_{\kappa,\lambda}(t,x)\dd x
=\int_{[0,\infty)}\kappa\psi(\tfrac{x}{\lambda})G(\kappa\lambda t,x)\dd x\text{ for all }t\geq0\text{ and }\psi\in C_c([0,\infty)).
\end{equation}
In order to obtain scaling solutions we thus require $\kappa\lambda\sim t^{-1}$, and since we are mostly interested in the form of solutions at large $x$ it makes sense to consider solutions for which there exist $p>0$ and a finite time-independent measure $\Phi\geq0$ such that
\begin{equation}\nonumber
\int_{(0,\infty)}\psi(x)G(t,x)\dd x=\int_{(0,\infty)}(t+1)^{\frac1p-1}\psi\big((t+1)^{\frac1p}x\big)\Phi(x)\dd x\text{ for all }t\geq0\text{ and }\psi\in C_c([0,\infty)).
\end{equation}
From mass conservation it then follows that for such solutions there should formally hold
\begin{equation}\nonumber
\int_{\{0\}}\psi(x)G(t,x)\dd x=\psi(0)\left(\int_{[0,\infty)}G(0,x)\dd x-(t+1)^{\frac1p-1}\int_{(0,\infty)}\Phi(x)\dd x\right),
\end{equation}
which by strict monotonicity of the origin implies that we require $p>1$. Solutions of this form with finite energy, i.e.~with $p=2$, were constructed in the second part of \cite{KV15}. In \cite{KV16} the construction was extended to the following
\begin{pr}\label{pr:2}
Given $\rho\in(1,2]$, there exists at least one nonnegative function $\Phi\in L^1(0,\infty)$ that for all $\psi\in C_c^1([0,\infty))$ satisfies
\begin{equation}\tag*{{\rm(SSPE)}$_\rho^{\rm w}$}\label{eq!:sspe_w}
\frac1\rho\int_{(0,\infty)}\big(x\psi'(x)-(\rho-1)(\psi(x)-\psi(0))\big)\Phi(x)\dd x
=\iint_{\{x>y>0\}}\frac{\Phi(x)\Phi(y)}{\sqrt{xy}}\Delta_y^2\psi(x)\dd y\dd x,
\end{equation}
and if $m(t)=M-(t+1)^{\frac1\rho-1}\|\Phi\|_{L^1(0,\infty)}\geq0$, and if $h(t,x)=(t+1)^{-1}\Phi((t+1)^{-\frac1\rho}x)$ defines a measure with density, then $G(t,\cdot)=m(t)\delta_0+h(t,\cdot)$ is a weak solution to \eqref{eq:qwte}.
\end{pr}
The proof of Proposition \ref{pr:2} comprised two steps. First measure-valued solutions were constructed, which were then shown to be sufficiently regular. Moreover, the regularity result provides local $\alpha$-H\"older regularity on $(0,\infty)$ with $\alpha<\frac12$, which in turn allows a bootstrap argument to\linebreak[1] show that solutions $\Phi$ to \ref{eq!:sspe_w} are actually smooth classical solutions to
\begin{multline}\tag*{{\rm(SSPE)}$_\rho$}\label{eq!:sspe}
-\tfrac1\rho x\Phi'(x)-\Phi(x)
=\int_0^{x/2}\left[\frac{\Phi(x+y)}{\sqrt{x+y}}+\frac{\Phi(x-y)}{\sqrt{x-y}}-2\frac{\Phi(x)}{\sqrt{x}}\right]\frac{\Phi(y)}{\sqrt{y}}\dd y\\
-2\frac{\Phi(x)}{\sqrt{x}}\int_{x/2}^x\frac{\Phi(y)}{\sqrt{y}}\dd y+\int_{x/2}^\infty\frac{\Phi(x+y)\Phi(y)}{\sqrt{(x+y)y}}\dd y.
\end{multline}
Several additional properties of solutions $\Phi$ to \ref{eq!:sspe} were proved in \cite{KV16}.

For $\rho\in(1,2)$ it was shown that any solution $\Phi$ to \ref{eq!:sspe} satisfies
\begin{multline}\label{eq:powerlawdecay}
\Phi(z)\sim\left.(2-\rho)(\rho-1)\|\Phi\|_\rho\right.z^{-\rho}\text{ as }z\rightarrow\infty,\\
\text{with }\|\Phi\|_\rho=\sup_{R>0}\textstyle\left\{R^{\rho-2}\int_{(0,\infty)}(x\wedge R)\Phi(x)\dd x\right\}<\infty.
\end{multline}
These so called fat-tailed solutions have finite mass, but their energies are infinite, which makes conservation of energy a void concept. However, for a solution $\Phi$ to \ref{eq!:sspe} the supremum in the definition of $\|\Phi\|_\rho$ coincides with the limit of the functional as $R\rightarrow\infty$, and for any self-similar solution $G$ to \eqref{eq:qwte} as constructed in Proposition \ref{pr:2} the norm $\|\cdot\|_\rho$ is constant. It may well be that the rate of divergence of the energy is conserved, as it is the case for specific higher moments in Smoluchowski's coagulation equation with solvable kernels (cf.~\cite{MP04}).

Now, if we suppose that $\Phi\geq0$ satisfies \ref{eq!:sspe_w} for all $\psi\in C_c^1([0,\infty))$ and with $\rho\geq2$, it can be shown that
\begin{equation}\nonumber
\textstyle\int_{(0,R)}x\Phi(x)\dd x\leq\rho\left(\int_{(0,R)}\Phi(x)\dd x\right)^2\leq\rho\|\Phi\|_{L^1(0,\infty)}^2\text{ for all }R>0.
\end{equation}
Consequently these profiles have finite energy, which for $\rho>2$ leads to the violation of conservation of energy by $G$. This is the reason for the restriction of the range of $\rho$ in Proposition \ref{pr:2}. For $\rho=2$, similar arguments as used to obtain the estimate above, yield bounds on all higher moments of $\Phi$. Using those bounds, it was shown that there exists a constant $a>0$ such that
\begin{equation}\nonumber
\Phi(z)\leq e^{-a z}\text{ for all }z\geq1.
\end{equation}
An exponential lower bound has thus far only been established in an integral sense.

\section{Two conjectures}
The approach taken in \cite{KV16} to prove pointwise exponential upper bounds on solutions $\Phi$ to \ref{eq!:sspe} with $\rho=2$, which consists of the use of explicit bounds on their higher moments, derives from \cite{NV14}. In light of the structural similarity between coagulation equations and \eqref{eq:qwte}, one would think many other results in \cite{NV14} could also be carried over. In particular, a pointwise exponential lower bound, and existence of the limit $\lim_{z\rightarrow\infty}-\frac1z\log(\Phi(z))$ are expected. However, it was noted in \cite{KV16} that a better understanding of the behaviour of the solution near zero seems to be required, even to be able to prove the lower bound.

Presently, it is only known that for any solution $\Phi$ to \ref{eq!:sspe} there holds
\begin{equation}\nonumber
\sup_{R>0}\textstyle\left\{\frac1{\sqrt{R}}\int_{(0,R)}\Phi(x)\dd x\right\}<\infty,
\end{equation}
but since we expect the solution to be well-behaved near zero, we pose the following
\begin{con}\label{con:3}
Given a solution $\Phi\in C_+^\infty((0,\infty))\cap L^1(0,\infty)$ to \ref{eq!:sspe} with $\rho\in(1,2]$, there holds
\begin{equation}\nonumber
\Phi(z)\sim\frac{A}{\sqrt{z}}\text{ as }z\rightarrow0,\text{ with }A=\sqrt{\tfrac6{\pi^2}\tfrac2\rho(\rho-1)\|\Phi\|_{L^1(0,\infty)}}.
\end{equation}
\end{con}\noindent
It is clear that this result holds if and only if
\begin{equation}\nonumber
\lim_{\lambda\rightarrow0^+}f_\lambda(x)=\frac{A}{x}\text{ for all }x>0,\text{ where }f_\lambda(x)=\lambda\frac{\Phi(\lambda x)}{\sqrt{\lambda x}},
\end{equation}
which, unfortunately, is still an open question. Using the weak formulation \ref{eq!:sspe_w}, we do have
\begin{equation}\nonumber
\lim_{\lambda\rightarrow0^+}\iint_{\{x>y>0\}}f_\lambda(x)f_\lambda(y)\Delta_y^2\psi(x)\dd x\dd y=\tfrac1\rho(\rho-1)\|\Phi\|_{L^1(0,\infty)}\times\psi(0)\text{ for all }\psi\in C_c^1([0,\infty)),
\end{equation}
and from this it can be found, either by writing the left hand side as a tested distributional second derivative, or by simply using $\psi(x)=(z-x)_+$, that
\begin{equation}\nonumber
\lim_{\lambda\rightarrow0^+}\iint_{\mathbb{R}_+^2}f_\lambda(x)f_\lambda(y)\big[(x+y-z)\wedge(z-|x-y|)\big]_+\dd x\dd y=A^2\tfrac{\pi^2}6\,z\text{ for all }z>0,
\end{equation}
which is consistent with the conjecture once we observe that
\begin{equation}\nonumber
\frac1z\iint_{\mathbb{R}_+^2}\frac{1}{xy}\big[(x+y-z)\wedge(z-|x-y|)\big]_+\dd x\dd y=\frac{\pi^2}6\text{ for all }z>0.
\end{equation}

Let us return to the likely decay behaviour of a solution $\Phi$ to \ref{eq!:sspe} with $\rho=2$. To that end we suppose that the limit $\lim_{z\rightarrow\infty}-\frac1z\log(\Phi(z))=:a>0$ exists, and that there exist constants $C>0$ and $\alpha\in\mathbb{R}$ such that $\Phi(z)\sim C\times z^\alpha e^{-az}$ as $z\rightarrow\infty$. Substituting this asymptotic behaviour for the tail of $\Phi$, we find that
\begin{equation}\nonumber
-\tfrac12 x\Phi'(x)-\Phi(x)\sim\tfrac12aC\times x^{\alpha+1}e^{-ax}\text{ as }x\rightarrow\infty,
\end{equation}
while for arbitrarily fixed $c\gg1$ we have
\begin{equation}\nonumber
\int_c^{x/2}\frac{\Phi(x-y)\Phi(y)}{\sqrt{(x-y)y}}\dd y\sim\left({\textstyle\int_{c/x}^{1/2}((1-z)z)^{\alpha-\frac12}\dd z}\right)C^2\times x^{2\alpha}e^{-ax}\text{ as }x\rightarrow\infty,
\end{equation}
where the integral between brackets converges if $\alpha>-\frac12$. Noting then that
\begin{multline}\nonumber
\left|\frac1{x^{\alpha-\frac12}e^{-ax}}\left[(x+y)^{\alpha-\frac12}e^{-a(x+y)}+(x-y)^{\alpha-\frac12}e^{-a(x-y)}-2x^{\alpha-\frac12}e^{-ax}\right]\right|\\
\leq4\sinh^2\big(\tfrac{ay}{2}\big)+O(\tfrac{y}{x})\text{ as }\tfrac{y}{x}\rightarrow\infty,
\end{multline}
we further obtain
\begin{equation}\nonumber
\int_0^c\left[\frac{\Phi(x+y)}{\sqrt{x+y}}+\frac{\Phi(x-y)}{\sqrt{x-y}}-2\frac{\Phi(x)}{\sqrt{x}}\right]\frac{\Phi(y)}{\sqrt{y}}\dd y\sim\left({\textstyle\int_0^c4\sinh^2(\frac{ay}{2})\frac{\Phi(y)}{\sqrt{y}}\dd y}\right)C\times x^{\alpha-\frac12}e^{-ax},
\end{equation}
where the integral between brackets is finite, and observing also that
\begin{equation}\nonumber
\int_c^x\frac{\Phi(x)\Phi(y)}{\sqrt{xy}}\dd y\sim\left({\textstyle\int_c^\infty y^{\alpha-\frac12}e^{-ay}\dd y}\right)C^2\times x^{\alpha-\frac12}e^{-ax},
\end{equation}
and
\begin{equation}\nonumber
\int_c^\infty\frac{\Phi(x+y)\Phi(y)}{\sqrt{(x+y)y}}\dd y\sim C^2\times \omega(x)e^{-ax},
\text{ with }\omega(x)={\textstyle\int_c^\infty((x+y)y)^{\alpha-\frac12}e^{-2ay}\dd y}=O(x^{\alpha-\frac12}),
\end{equation}
we arrive by matching of asymptotics at the following
\begin{con}\label{con:4}
Given a solution $\Phi\in C_+^\infty((0,\infty))\cap L^1(0,\infty)$ to \ref{eq!:sspe} with $\rho=2$, there exists a constant $a>0$ such that
\begin{equation}\label{eq:con_exp}
\Phi(z)\sim \tfrac8\pi aze^{-az}\text{ as }z\rightarrow\infty.
\end{equation}
\end{con}\noindent
In order to further support the claims of Conjectures \ref{con:3} and \ref{con:4}, we have implemented a numerical scheme to compute solutions to \ref{eq!:sspe}. This was achieved by a finite element approximation with base functions $(x_n-x)_+$, $(x_n)\in \mathbb{R}_+^N$, which was then solved by Newton's method.
%\documentclass[a4paper,10pt]{article}
%\usepackage{tikz}
%	\usetikzlibrary{decorations.pathreplacing}
%\usepackage{pgfplots}
%\usepackage[a4paper,top=1in, bottom=1.25in, left=1.2in, right=1.2in]{geometry}
%\begin{document}

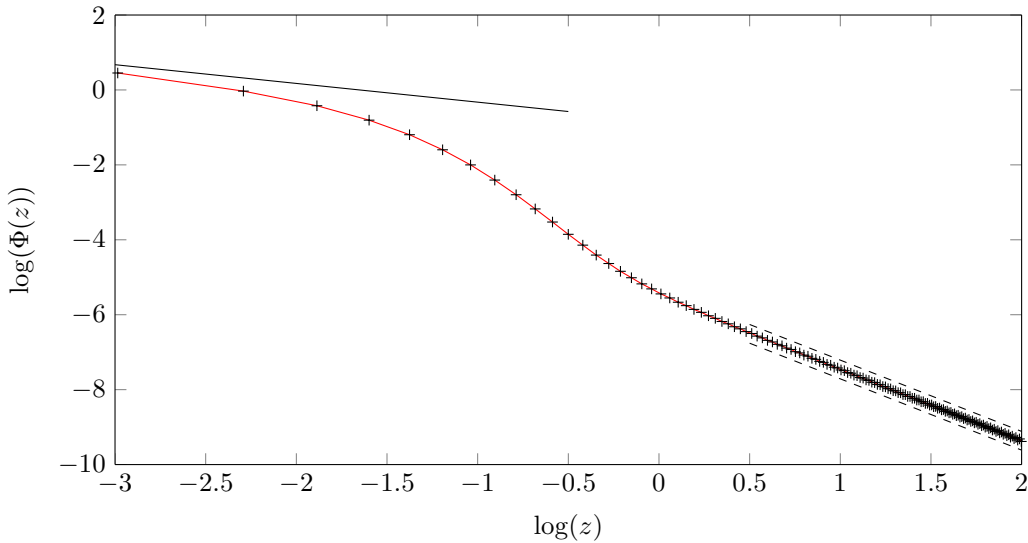
\begin{figure}[h]\center
\begin{tikzpicture}
\begin{axis}[%
width=.8\textwidth,
height=.4\textwidth,
scale only axis,
separate axis lines,
every outer x axis line/.append style={black},
every x tick label/.append style={font=\color{black}},
xmin=-3,
xmax=2,
every outer y axis line/.append style={black},
every y tick label/.append style={font=\color{black}},
ymin=-10,
ymax=2,
axis background/.style={fill=white},
xlabel=$\log(z)$,
ylabel=$\log(\Phi(z))$
]
\addplot [color=red,forget plot]
  table[row sep=crcr]{%
-2.98509925595043	0.452640320154833\\
-2.29195207539048	-0.0310391531201736\\
-1.88648696728232	-0.421440484192858\\
-1.59880489483054	-0.806554196378371\\
-1.37566134351633	-1.19517137364318\\
-1.19333978672237	-1.59731728610511\\
-1.03918910689511	-2.00076738090428\\
-0.905657714270593	-2.40577906429791\\
-0.787874678614209	-2.79697263287336\\
-0.682514162956383	-3.17575522172836\\
-0.587203983152058	-3.5257114964602\\
-0.500192606162428	-3.85350374155841\\
-0.420149898488892	-4.14267369619307\\
-0.34604192633517	-4.40802776067126\\
-0.277049054848218	-4.63320623057931\\
-0.212510533710647	-4.84081850908948\\
-0.151885911894212	-5.01275816975035\\
-0.0947274980542638	-5.17648013640991\\
-0.0406602767839879	-5.31030241935341\\
0.0106330176035627	-5.44410580724806\\
0.0594231817729945	-5.55230785130702\\
0.105943197407887	-5.66631137479372\\
0.150394959978721	-5.75718457944826\\
0.192954574397517	-5.85775470761439\\
0.233776568917772	-5.93635348094816\\
0.272997282071054	-6.02734299414626\\
0.310737610053901	-6.09672752730665\\
0.347105254224776	-6.18051671728365\\
0.382196574036046	-6.24261009284221\\
0.416098125711727	-6.32074686824649\\
0.448887948534718	-6.37682777799708\\
0.480636646849298	-6.4503811725753\\
0.511408305516052	-6.50135065225147\\
0.541261268665733	-6.57110106702962\\
0.570248805538985	-6.61762664619677\\
0.598419682505682	-6.68417008137988\\
0.625818656693796	-6.72676618960605\\
0.652486903775957	-6.79057386899057\\
0.678462390179218	-6.82964897816066\\
0.703780198163508	-6.89110338416834\\
0.728472810753879	-6.92698924901703\\
0.75257036233294	-6.98640716453138\\
0.776100859743134	-7.01937796748294\\
0.799090377967833	-7.07702598196991\\
0.821563233819891	-7.10731147347791\\
0.843542140538667	-7.16341687489555\\
0.86504834575963	-7.19121174915365\\
0.886101754957463	-7.24597044535497\\
0.906721042160198	-7.27144124224543\\
0.926923749477718	-7.32502368366443\\
0.946726376773897	-7.34831399885908\\
0.966144462630999	-7.40086970712795\\
0.985192657601694	-7.42210420617709\\
1.00388479061385	-7.47376530114275\\
1.02223392928204	-7.49305286491256\\
1.04025243478472	-7.54393685417885\\
1.05795201188412	-7.56137307938639\\
1.07534375459599	-7.61158509406452\\
1.09243818795529	-7.62725430567727\\
1.10924530627167	-7.67688891247128\\
1.12577460822288	-7.69086580130557\\
1.14203512909466	-7.74000848666368\\
1.1580354704411	-7.7523594544942\\
1.17378382740924	-7.80108785071391\\
1.18928801394521	-7.81187212498991\\
1.204555486076	-7.86025703161004\\
1.21959336344054	-7.86952759614914\\
1.23440844922568	-7.91763383710077\\
1.24900724864683	-7.92543821513169\\
1.26339598609893	-7.97332536294155\\
1.27758062109089	-7.97970627936886\\
1.29156686306563	-8.02742927021799\\
1.30536018519796	-8.03242521669042\\
1.31896583725374	-8.08003487686735\\
1.33238885758588	-8.0836805937107\\
1.3456340843359	-8.13122409221184\\
1.35870616590326	-8.13355098491898\\
1.37160957073916	-8.18107222324704\\
1.38434859651659	-8.18210872171742\\
1.39692737872345	-8.22964867318357\\
1.40934989872201	-8.22942054394636\\
1.42161999131382	-8.27701754855546\\
1.43374135184617	-8.27554816715643\\
1.44571754289289	-8.32323818959613\\
1.45755200053989	-8.32054877889683\\
1.46924804030308	-8.36836563577431\\
1.48080886270416	-8.36447547492603\\
1.49223755852778	-8.4124510344761\\
1.50353711378171	-8.40737764438726\\
1.51471041437984	-8.45554200212317\\
1.52576025056642	-8.44930130850924\\
1.53668932109861	-8.49768294459702\\
1.54750023720283	-8.49028942401341\\
1.55819552631958	-8.53891534141319\\
1.56877763565011	-8.53038214873346\\
1.57924893551741	-8.57927800050477\\
1.58961172255295	-8.56961708463367\\
1.59986822272014	-8.61880728407698\\
1.61002059418416	-8.60802949374929\\
1.62007093003766	-8.65753732482656\\
1.63002126089083	-8.64565257647798\\
1.639971591744	-8.69569014174349\\
1.64992192259717	-8.6830670386055\\
1.65987225345034	-8.73384463802105\\
1.6698225843035	-8.72045871065265\\
1.67977291515667	-8.77200116709584\\
1.68972324600984	-8.75782719173608\\
1.69967357686301	-8.81016016757951\\
1.70962390771618	-8.79517206203493\\
1.71957423856934	-8.84832208073789\\
1.72952456942251	-8.83249286905165\\
1.73947490027568	-8.88648734692731\\
1.74942523112885	-8.86978913130066\\
1.75937556198202	-8.92465640559422\\
1.76932589283518	-8.90706033191142\\
1.77927622368835	-8.96282969636674\\
1.78922655454152	-8.94430592369286\\
1.79917688539469	-9.00100765756552\\
1.80912721624786	-8.98152532382034\\
1.81907754710103	-9.03919073251574\\
1.82902787795419	-9.01871791564414\\
1.83897820880736	-9.07737936433965\\
1.84892853966053	-9.05588304637959\\
1.8588788705137	-9.11557400224775\\
1.86882920136687	-9.093020028486\\
1.87877953222003	-9.1537750949106\\
1.8887298630732	-9.13012814098951\\
1.89868019392637	-9.19198310006825\\
1.90863052477954	-9.16720662093592\\
1.91858085563271	-9.23019847975281\\
1.92853118648587	-9.20425467224631\\
1.93848151733904	-9.26842170006652\\
1.94843184819221	-9.24127146013776\\
1.95838217904538	-9.30665323652035\\
1.96833250989855	-9.27825611085818\\
1.97828284075171	-9.34489356974101\\
1.98823317160488	-9.31520771143021\\
1.99818350245805	-9.38314319330481\\
};
\addplot [color=black,only marks,mark=+,mark options={solid},forget plot]
  table[row sep=crcr]{%
-2.98509925595043	0.452640320154833\\
-2.29195207539048	-0.0310391531201736\\
-1.88648696728232	-0.421440484192858\\
-1.59880489483054	-0.806554196378371\\
-1.37566134351633	-1.19517137364318\\
-1.19333978672237	-1.59731728610511\\
-1.03918910689511	-2.00076738090428\\
-0.905657714270593	-2.40577906429791\\
-0.787874678614209	-2.79697263287336\\
-0.682514162956383	-3.17575522172836\\
-0.587203983152058	-3.5257114964602\\
-0.500192606162428	-3.85350374155841\\
-0.420149898488892	-4.14267369619307\\
-0.34604192633517	-4.40802776067126\\
-0.277049054848218	-4.63320623057931\\
-0.212510533710647	-4.84081850908948\\
-0.151885911894212	-5.01275816975035\\
-0.0947274980542638	-5.17648013640991\\
-0.0406602767839879	-5.31030241935341\\
0.0106330176035627	-5.44410580724806\\
0.0594231817729945	-5.55230785130702\\
0.105943197407887	-5.66631137479372\\
0.150394959978721	-5.75718457944826\\
0.192954574397517	-5.85775470761439\\
0.233776568917772	-5.93635348094816\\
0.272997282071054	-6.02734299414626\\
0.310737610053901	-6.09672752730665\\
0.347105254224776	-6.18051671728365\\
0.382196574036046	-6.24261009284221\\
0.416098125711727	-6.32074686824649\\
0.448887948534718	-6.37682777799708\\
0.480636646849298	-6.4503811725753\\
0.511408305516052	-6.50135065225147\\
0.541261268665733	-6.57110106702962\\
0.570248805538985	-6.61762664619677\\
0.598419682505682	-6.68417008137988\\
0.625818656693796	-6.72676618960605\\
0.652486903775957	-6.79057386899057\\
0.678462390179218	-6.82964897816066\\
0.703780198163508	-6.89110338416834\\
0.728472810753879	-6.92698924901703\\
0.75257036233294	-6.98640716453138\\
0.776100859743134	-7.01937796748294\\
0.799090377967833	-7.07702598196991\\
0.821563233819891	-7.10731147347791\\
0.843542140538667	-7.16341687489555\\
0.86504834575963	-7.19121174915365\\
0.886101754957463	-7.24597044535497\\
0.906721042160198	-7.27144124224543\\
0.926923749477718	-7.32502368366443\\
0.946726376773897	-7.34831399885908\\
0.966144462630999	-7.40086970712795\\
0.985192657601694	-7.42210420617709\\
1.00388479061385	-7.47376530114275\\
1.02223392928204	-7.49305286491256\\
1.04025243478472	-7.54393685417885\\
1.05795201188412	-7.56137307938639\\
1.07534375459599	-7.61158509406452\\
1.09243818795529	-7.62725430567727\\
1.10924530627167	-7.67688891247128\\
1.12577460822288	-7.69086580130557\\
1.14203512909466	-7.74000848666368\\
1.1580354704411	-7.7523594544942\\
1.17378382740924	-7.80108785071391\\
1.18928801394521	-7.81187212498991\\
1.204555486076	-7.86025703161004\\
1.21959336344054	-7.86952759614914\\
1.23440844922568	-7.91763383710077\\
1.24900724864683	-7.92543821513169\\
1.26339598609893	-7.97332536294155\\
1.27758062109089	-7.97970627936886\\
1.29156686306563	-8.02742927021799\\
1.30536018519796	-8.03242521669042\\
1.31896583725374	-8.08003487686735\\
1.33238885758588	-8.0836805937107\\
1.3456340843359	-8.13122409221184\\
1.35870616590326	-8.13355098491898\\
1.37160957073916	-8.18107222324704\\
1.38434859651659	-8.18210872171742\\
1.39692737872345	-8.22964867318357\\
1.40934989872201	-8.22942054394636\\
1.42161999131382	-8.27701754855546\\
1.43374135184617	-8.27554816715643\\
1.44571754289289	-8.32323818959613\\
1.45755200053989	-8.32054877889683\\
1.46924804030308	-8.36836563577431\\
1.48080886270416	-8.36447547492603\\
1.49223755852778	-8.4124510344761\\
1.50353711378171	-8.40737764438726\\
1.51471041437984	-8.45554200212317\\
1.52576025056642	-8.44930130850924\\
1.53668932109861	-8.49768294459702\\
1.54750023720283	-8.49028942401341\\
1.55819552631958	-8.53891534141319\\
1.56877763565011	-8.53038214873346\\
1.57924893551741	-8.57927800050477\\
1.58961172255295	-8.56961708463367\\
1.59986822272014	-8.61880728407698\\
1.61002059418416	-8.60802949374929\\
1.62007093003766	-8.65753732482656\\
1.63002126089083	-8.64565257647798\\
1.639971591744	-8.69569014174349\\
1.64992192259717	-8.6830670386055\\
1.65987225345034	-8.73384463802105\\
1.6698225843035	-8.72045871065265\\
1.67977291515667	-8.77200116709584\\
1.68972324600984	-8.75782719173608\\
1.69967357686301	-8.81016016757951\\
1.70962390771618	-8.79517206203493\\
1.71957423856934	-8.84832208073789\\
1.72952456942251	-8.83249286905165\\
1.73947490027568	-8.88648734692731\\
1.74942523112885	-8.86978913130066\\
1.75937556198202	-8.92465640559422\\
1.76932589283518	-8.90706033191142\\
1.77927622368835	-8.96282969636674\\
1.78922655454152	-8.94430592369286\\
1.79917688539469	-9.00100765756552\\
1.80912721624786	-8.98152532382034\\
1.81907754710103	-9.03919073251574\\
1.82902787795419	-9.01871791564414\\
1.83897820880736	-9.07737936433965\\
1.84892853966053	-9.05588304637959\\
1.8588788705137	-9.11557400224775\\
1.86882920136687	-9.093020028486\\
1.87877953222003	-9.1537750949106\\
1.8887298630732	-9.13012814098951\\
1.89868019392637	-9.19198310006825\\
1.90863052477954	-9.16720662093592\\
1.91858085563271	-9.23019847975281\\
1.92853118648587	-9.20425467224631\\
1.93848151733904	-9.26842170006652\\
1.94843184819221	-9.24127146013776\\
1.95838217904538	-9.30665323652035\\
1.96833250989855	-9.27825611085818\\
1.97828284075171	-9.34489356974101\\
1.98823317160488	-9.31520771143021\\
1.99818350245805	-9.38314319330481\\
};
\addplot [color=black,solid,forget plot]
  table[row sep=crcr]{%
-3	0.674116238129489\\
-2.5	0.424116238129489\\
-2	0.174116238129489\\
-1.5	-0.0758837618705106\\
-1	-0.325883761870511\\
-0.5	-0.575883761870511\\
};
\addplot [color=black,dashed,forget plot]
  table[row sep=crcr]{%
0.5	-6.26010852694256\\
1	-7.21155319448438\\
1.5	-8.16299786202619\\
2	-9.11444252956801\\
};
\addplot [color=black,dashed,forget plot]
  table[row sep=crcr]{%
0.5	-6.76010852694256\\
1	-7.71155319448438\\
1.5	-8.66299786202619\\
2	-9.61444252956801\\
};
\end{axis}
\end{tikzpicture}
\caption{Numerical computation of a solution $\Phi$ to \ref{eq!:sspe} for $\rho=1.9$. The solid line corresponds to the conjectured asymptitic behaviour near the origin (cf.~Conjecture \ref{con:3}). The dashed lines have slope $-\rho$, indicating agreement between numerics and the theoretical decay [cf.~\eqref{eq:powerlawdecay}].}\label{fig:ssprho19}
\end{figure}

%\end{document}
\input{selfsimprofrho20.tex}
\newpage
Lastly, we stress that Conjectures \ref{con:3} and \ref{con:4} are consistent with the following scaling property.
\begin{lm}\label{lm:5}
Every solution $\Phi\in C_+^\infty((0,\infty))\cap L^1(0,\infty)$ to \ref{eq!:sspe} gives rise to a one-parameter family $\{\Phi_c\}_{c>0}$ of rescaled solutions to \ref{eq!:sspe}, given by $\Phi_c(x)=\Phi(cx)$ for all $c,x>0$.
\end{lm}\noindent
In particular, if solutions to \ref{eq!:sspe} were shown to be unique up to rescaling, then Conjecture \ref{con:4} would be equivalent to existence of a constant $\hat{a}>0$ such that \eqref{eq:con_exp} holds with $a=\hat{a}/\|\Phi\|_{L^1(0,\infty)}$. As the integral of an exponential tail is negligible, it is actually possible for us to see such a relation.% (cf.~Figure \ref{fig:avsint}).
%\documentclass[a4paper,10pt]{article}
%\usepackage{tikz}
%	\usetikzlibrary{decorations.pathreplacing}
%\usepackage{pgfplots}
%\usepackage[a4paper,top=1in, bottom=1.25in, left=1.2in, right=1.2in]{geometry}
%\begin{document}

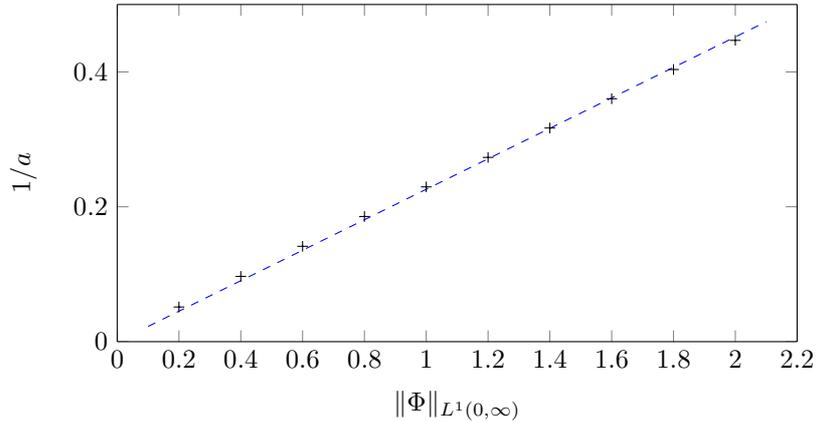
\begin{figure}[h]\center
\begin{tikzpicture}

\begin{axis}[%
width=.6\textwidth,
height=.3\textwidth,
scale only axis,
separate axis lines,
every outer x axis line/.append style={black},
every x tick label/.append style={font=\color{black}},
xmin=0,
xmax=2.2,
every outer y axis line/.append style={black},
every y tick label/.append style={font=\color{black}},
ymin=0,
ymax=0.5,
axis background/.style={fill=white},
xlabel=$\|\Phi\|_{L^1(0,\infty)}$,
ylabel=$1/a$
]
\addplot [color=black,only marks,mark=+,mark options={solid},forget plot]
  table[row sep=crcr]{%
0.2	0.0514154889727612\\
0.4	0.0968076060768352\\
0.6	0.141422172167861\\
0.8	0.185617210771189\\
1	0.229538527542913\\
1.2	0.273260994519005\\
1.4	0.316828587951689\\
1.6	0.36026940920181\\
1.8	0.403602511902056\\
2	0.446841792082628\\
};
\addplot [color=blue,dashed,forget plot]
  table[row sep=crcr]{%
0.1	0.0225971695880058\\
1.1	0.248568865468063\\
2.1	0.474540561348121\\
};
\end{axis}
\end{tikzpicture}
\caption{For numerically computed solutions $\Phi$ to \ref{eq!:sspe} with $\rho=2$, we have determined $a$ from a least squares fit to the conjectured tail behaviour. As expected, we observe an inversely proportional relation between $a$ and $\|\Phi\|_{L^1(0,\infty)}$.}\label{fig:avsint}
\end{figure}

%\end{document}

\newpage\appendix
\linespread{1.2}
\section{Appendix}
For the sake of completeness we include the lemma below. Its proof is usually omitted, being described as ``some algebra and taking into account the energy conservation'' in \cite{ST97}.
\begin{lm}\label{lm:6}
Given $k_1,k_2,k_3,k_4\geq0$ such that $k_1^2+k_2^2=k_3^2+k_4^2$, then there holds
\begin{equation}\label{eq:lm_6}
\int_{\mathbb{R}_+}\sin(k_1s)\sin(k_2s)\sin(k_3s)\sin(k_4s)s^{-2}\dd s=\tfrac\pi4\min\{k_1,k_2,k_3,k_4\}.
\end{equation}
\end{lm}
\begin{proof}
Supposing that $k_1\geq k_3\geq k_4\geq k_2$ ({\sc wlog}), and observing that the integrand on the left hand side is symmetric, then \eqref{eq:lm_6} reduces to
\begin{equation}\label{eq:lm_6_1}
\int_\mathbb{R}\sin(k_1s)\sin(k_2s)\sin(k_3s)\sin(k_4s)s^{-2}\dd s=\tfrac\pi2k_2.
\end{equation}
For any $s\in\mathbb{R}$, one can further check that
\begin{multline}\nonumber\textstyle
\sin(k_1s)\sin(k_2s)\sin(k_3s)\sin(k_4s)
=\frac1{16}\prod_{\ell=1}^4(e^{ik_\ell s}-e^{-ik_\ell s})\\\textstyle
=\frac1{16}\sum_{j_1,j_2,j_3,j_4=1}^2\prod_{\ell=1}^4(-1)^{j_\ell}e^{i(-1)^{j_\ell}k_\ell s},
\end{multline}
so the left hand side of \eqref{eq:lm_6_1} can be written as
\begin{equation}\label{eq:lm_6_3}
\frac1{16}\sum_{j_1,j_2,j_3,j_4=1}^2(-1)^{j_1+j_2+j_3+j_4}\int_\mathbb{R}\exp\Big\{i\Big(\textstyle\sum_{\ell=1}^4(-1)^{j_\ell}k_\ell\Big)s\Big\}s^{-2}\dd s.
\end{equation}
Computing the integrals by standard methods we then find that \eqref{eq:lm_6_3} equals
\begin{multline}\label{eq:lm_6_4}
\frac\pi{16}\sum_{j_1,j_2,j_3,j_4=1}^2(-1)^{j_1+j_2+j_3+j_4+1}\Big|\textstyle\sum_{\ell=1}^4(-1)^{j_\ell}k_\ell\Big|\\
=\frac\pi{16}\sum_{j_1=1}^2\sum_{j_2,j_3,j_4=1}^2(-1)^{(j_2+j_1)+(j_3+j_1)+(j_4+j_1)+1}\Big|k_1+\textstyle\sum_{\ell=2}^4(-1)^{j_\ell+j_1}k_\ell\Big|\\
=\frac\pi8\sum_{j_2',j_3',j_4'=1}^2(-1)^{j_2'+j_3'+j_4'+1}\Big|k_1+\textstyle\sum_{\ell=2}^4(-1)^{j_\ell'}k_\ell\Big|.
\end{multline}
By our assumptions that $k_1\geq k_3\geq k_4\geq k_2$, it now follows that
\begin{equation}\nonumber
-|k_1+k_2+k_3+k_4|+|k_1+k_2+k_3-k_4|+|k_1+k_2-k_3+k_4|+|k_1-k_2+k_3+k_4|=2k_1,
\end{equation}
hence the sum in the right hand side of \eqref{eq:lm_6_4} is given by
\begin{multline}\nonumber%\label{eq:lm_6_6}
|k_1-k_2-k_3-k_4|+2k_1-|k_1-k_2+k_3-k_4|-|k_1+k_2-k_3-k_4|-|k_1-k_2-k_3+k_4|\\
=|k_1-k_3-(k_4+k_2)|+(k_1-k_3+k_4+k_2)-|k_1-k_3-(k_4-k_2)|-|k_1-k_3+(k_4-k_2)|\\
=2\big((k_1-k_3)\vee(k_4+k_2)\big)-2\big((k_1-k_3)\vee(k_4-k_2)\big).
\end{multline}
Using lastly the identity $k_1^2+k_2^2=k_3^2+k_4^2$ to find that the maxima are $(k_4+k_2)$ and $(k_4-k_2)$ respectively, the lemma follows.
\end{proof}

\newpage\linespread{1}
\begin{ack}
Sincere thanks to M.~Rumpf for his advise on the numerical computation of self-similar profiles for \eqref{eq:qwte}.

The author is supported through {\em CRC 1060 The mathematics of emergent effects} at the University of Bonn, that is funded by the German Science Foundation (DFG).
\end{ack}

\end{document}